\documentclass[%
reprint,  
amsmath,amssymb,
aps,
prd,
]{revtex4-2}

\usepackage{graphicx}
\usepackage{ctable}        
\usepackage{multirow}
\usepackage{array}         
\usepackage{xcolor}        
  \definecolor{navy}{HTML}{2c1fa5}
  \definecolor{maroon}{HTML}{800000}
  \definecolor{gray}{HTML}{808080}
\usepackage[final]{hyperref} 
\hypersetup{
	colorlinks=true,         
  linkcolor=navy,          
	citecolor=navy,          
	filecolor=magenta,       
	urlcolor=navy         
}
\usepackage{chngcntr}      

\usepackage{texmacros}     

\begin{document}

\title{Relativistic head-on collisions of {$U(1)$} gauged \qballs{}}

\author{Michael P. Kinach}
 \email{mikin@physics.ubc.ca}
\author{Matthew W. Choptuik}%
 \email{choptuik@physics.ubc.ca}
\affiliation{%
  Department of Physics and Astronomy, University of British Columbia,\\
  6224 Agricultural Road, Vancouver, British Columbia, V6T 1Z1, Canada
}%

\date{\today}

\begin{abstract}

  We investigate the collision dynamics of $U(1)$ gauged \qballs{} by
  performing high-resolution numerical simulations in axisymmetry.
  Focusing on the case of relativistic head-on collisions, we consider
  the effects of the initial velocity, relative phase, relative charge, and
  electromagnetic coupling strength on the outcome of the collision. We
  find that the collision dynamics can depend strongly on these
  parameters; most notably, electromagnetic effects can significantly
  alter the outcome of the collision when the gauge coupling is large.
  When the gauge coupling is small, we find that the dynamics generally
  resemble those of ordinary (non-gauged) \qballs{}. 

\end{abstract}

\maketitle


\section{\label{sec:intro} Introduction }

The study of non-linear wave equations has a long and rich history in modern
physics. One of the most remarkable insights to emerge from this tradition has
been the discovery of \textit{solitons}: localized solutions to the field
equations that can propagate without dispersing.  In many respects, solitons
behave like a rudimentary model of a particle which can be constructed from
smooth classical fields. They can generally be classified as either topological
or non-topological depending on whether the underlying model has a non-trivial
topology.  Examples of topological solitons include the kink/anti-kink
solutions of quantum field theory, skyrmions and vortices in condensed matter
physics, and cosmological domain walls \cite{Manton2004,Shnir2018}. In
contrast, non-topological solitons can arise due a balancing between the
effects of non-linearity and dispersion and are often characterized by the
existence of a conserved Noether charge \cite{Shnir2018}. The prototypical
examples of non-topological solitons are \textit{\qballs{}} which arise in
complex scalar field theories admitting a $U(1)$ symmetry.

The study of \qballs{} began in earnest with the work of Coleman \cite{Coleman1985,*Coleman1986} who
described them as localized solutions of a complex scalar field theory with a
non-linear attractive potential and a global $U(1)$ symmetry. This work has
since been extended to show that \qball{} solutions can arise in a variety of
physically-motivated models (see \cite{Nugaev2020} for a review). In the
context of cosmology and particle physics, \qballs{} may be relevant for
various early-Universe scenarios such as baryogenesis and the dark matter
problem \cite{Kasuya2000,Dine2003,Kusenko1998,Kusenko2001}. They may also arise
in the context of non-linear optics \cite{Radu2008} and condensed matter
systems \cite{Enqvist2003,Bunkov2007}.  Mathematically, \qballs{} are
characterized by the presence of a conserved Noether charge $Q$ which is
associated with the $U(1)$ symmetry of the theory.  The global $U(1)$ symmetry
can also be made into a local $U(1)$ symmetry via the introduction of a $U(1)$
gauge field; the resulting solutions are called \textit{gauged \qballs{}} and
represent a coupling of the system to electromagnetism \cite{Lee1989}.

While the basic properties of \qballs{} are well-known, it remains a
challenging problem to model their full time-dependent dynamical behaviour.
This is due mainly to the non-linear structure of the underlying equations
which typically requires a numerical treatment. Early work on this topic
revealed that \qball{} dynamics can be remarkably complex, particularly when
considering interactions and relativistic collisions of \qballs{}. Perhaps the
most comprehensive studies of this type were performed by Axenides \textit{et
al.}~in two spatial dimensions \cite{Axenides2000} and Battye and Sutcliffe in
three spatial dimensions \cite{Battye2000}. There it was shown that \qballs{}
can interact elastically or inelastically depending on the collision
parameters.  They may also transfer charge, annihilate, or form oscillatory
charge-swapping structures \cite{Copeland2014,Xie2021,Hou2022} under the right
conditions.  Additional studies have also considered different scalar field
models, higher collision velocities, or greater numerical resolutions
\cite{Multamaki2000a,Multamaki2000b,Al-alawi2009,Gutierrez2013,Dvali2024}.  A
general conclusion to be drawn from these studies is that \qball{} behaviour
can be quite complex and unexpected.

In the present paper, we continue this exploration of \qball{} dynamics by
considering relativistic head-on collisions of $U(1)$ gauged \qballs{} in
axisymmetry.  Intuitively, one might expect that the addition of the $U(1)$
gauge field may lead to novel dynamical behaviour due to the interaction of
electromagnetic charges and currents.  However, this possibility has remained
largely unexplored in the literature. 
Our aim is to shed light on this
topic by performing fully non-linear numerical evolutions of the field
equations in axisymmetry. We explore the effects of various collision
parameters such as the initial velocity, relative phase, relative charge, and
electromagnetic coupling strength in order to gain insight on the general
phenomenology of gauged \qball{} collisions.

In a previous paper \cite{Kinach2023}, we numerically investigated the
dynamical behaviour of $U(1)$ gauged \qballs{} when subject to axisymmetric
perturbations. There it was found that stable gauged \qball{} configurations
can exist in both logarithmic and polynomial models. Using these solutions as a
starting point, we construct binary gauged \qball{} initial data consisting of
two stable solutions which are boosted toward each other at relativistic
velocities. We then evolve the system according to the equations of motion and
observe the subsequent dynamics. 

When the gauge coupling is small, our results parallel those found for ordinary
(non-gauged) \qball{} collisions.  Specifically, we find that the collision
dynamics can be divided into three regimes---which we will call the
\textit{elastic}, \textit{fragmentation}, and \textit{merger} 
regimes---depending on the incident velocity of the colliding \qballs{}. In the elastic
regime (corresponding to high velocities), the collisions are primarily elastic
with the \qballs{} passing through each other virtually unscathed and forming a
destructive interference pattern at the moment of impact. In the merger and
fragmentation regimes (corresponding to low and intermediate velocities,
respectively), the collisions are primarily inelastic with several possible
outcomes. At the lowest velocities, the \qballs{} can merge into a single
\qball{} of a larger size, while at intermediate velocities they tend to
fragment into many pieces.  We also investigate collisions of
oppositely-charged and phase-shifted \qball{s}, finding evidence for
annihilation and charge transfer, respectively.

When the gauge coupling is large,
we find that electromagnetic effects can
significantly alter the outcome of the collision.  For gauged \qballs{} with
charge of equal sign, we find that the Coulomb repulsion tends to decelerate
the \qballs{} prior to the moment of impact. At low incident velocities, this
can prevent the interaction of the \qball{} fields entirely; at higher
velocities, it simply reduces the effective collision velocity. We also find
that collisions at large gauge coupling are rarely an elastic process. Unlike
the free-passage behaviour observed for small gauge coupling, 
the collision of gauged \qballs{} at high-velocities tends to result in the
formation of ring-like objects (which we have previously called ``gauged
\qrings" \cite{Kinach2023}) or elongated structures even for collision
velocities very close to the speed of light. For collisions involving
\qballs{} of unequal phase, we again observe charge transfer similar to the
case of small gauge coupling.
However, we find that the gauged \qball{}s created in this process often break
apart, presumably due to the reduced range of stable solutions which exist at
large gauge coupling. For collisions of oppositely-charged \qballs{}, the
Coulomb force accelerates the \qballs{} prior to the moment of impact. These
collisions can result in the annihilation of significant charge and the
production of an electromagnetic radiation pulse.  In sum, we find that the
collision of gauged \qballs{} can be a violent process with some striking
differences when compared to the non-gauged case.

The outline of this paper is as follows: in Sec.~\ref{sec:review}, we briefly
review the theory of $U(1)$ gauged \qballs{}.  In Sec.~\ref{sec:numerics}, we
summarize our numerical approach to the head-on collision problem.  In
Sec.~\ref{sec:results}, we present our main results and summarize the general
dynamics observed for $U(1)$ gauged \qball{} collisions.  In
Sec.~\ref{sec:conclusion}, we provide some concluding remarks.

In this work, we use units where $c=\hbar=1$ and employ the metric signature
$(-,+,+,+)$.  For brevity, we will interchangeably use the terms ``\qball{}"
and ``gauged \qball{}" when referring to \qballs{} coupled to the
electromagnetic field. When referring to \qballs{} which do not admit any such
coupling, we will explicitly use the term ``non-gauged \qball{}".

\section{\label{sec:review} Review of $U(1)$ Gauged \qballs{}}

For a system composed of a complex scalar field $\phi$ coupled to a $U(1)$
gauge field, $A_\mu$, the Lagrangian density takes the form
\begin{equation}
  \mathcal{L} = -\left(D_\mu\phi\right)^* D^\mu\phi-V\left(|\phi|\right)-\frac{1}{4}F_{\mu\nu}F^{\mu\nu}.
  \label{eqn:gauged-lagr}
\end{equation}
Here, $F_{\mu\nu}=\partial_{\mu} A_\nu - \partial_{\nu} A_\mu$ is the
electromagnetic field tensor, $D_\mu = \nabla_\mu-ieA_\mu$ describes the gauge
covariant derivative with coupling constant $e$, and $V(|\phi|)$ represents a
$U(1)$-invariant scalar field potential.
The equations of motion for the theory
\eqref{eqn:gauged-lagr} take the form
\begin{align}
  D_\mu D^\mu \phi - \frac{\partial}{\partial \phi^*}V(|\phi|)&=0, \label{eqn:eom-a}\\
  \nabla_\mu F^{\mu\nu}+ej^\nu&=0, \label{eqn:eom-b}
\end{align}
where $j^\nu$ is the Noether current density,
\begin{equation} \label{eqn:j-current}
  j^\nu=-i(\phi^* D^\nu\phi-\phi(D^\nu\phi)^*).
\end{equation}
This quantity can be integrated to obtain the conserved Noether charge
$Q=\int j^{0}\,d^3x$ associated with the $U(1)$ symmetry of the theory.
Likewise, there exists a conserved energy $E=\int T_{00}\,d^3x$
which can be computed from the energy-momentum tensor of the theory,
\begin{equation} \label{eqn:emt}
  \begin{split}
    T_{\mu\nu} =\, &F_{\mu\alpha} F_{\nu\beta} g^{\beta\alpha} -\frac{1}{4}g_{\mu\nu}F_{\alpha\beta}F^{\alpha\beta}\\
  &+D_\mu\phi(D_\nu\phi)^*+D_\nu\phi(D_\mu\phi)^*\\
  &-g_{\mu\nu}(D_\alpha\phi(D^\alpha\phi)^*+V(|\phi|)).
  \end{split}
\end{equation}

Solutions to the equations of motion \eqref{eqn:eom-a}--\eqref{eqn:eom-b} which
represent gauged \qballs{} can be found by making a spherically-symmetric
ansatz for the fields,
\begin{align}
  \phi(t,\vec{x})&=f(r)e^{i\omega t},  \label{eqn:sph-ansatz-f} \\
  A_0(t,\vec{x})&=A_0(r), \label{eqn:sph-ansatz-A} \\
  A_i(t,\vec{x})&=0,
\end{align}
and imposing the boundary conditions
\begin{alignat}{2}
  \lim_{r\rightarrow \infty}f(r)=0,\qquad\quad &&\frac{df}{dr}(0)=0,\\
  \lim_{r\rightarrow \infty}A_0(r)=0,\qquad\quad && \frac{dA_0}{dr}(0)=0.
\end{alignat}
This ansatz yields the reduced equations of motion
\begin{align}
  f''(r)+\frac{2}{r}f'(r)+f(r)g(r)^2-\frac{1}{2}\frac{d}{df}V(f)&=0,
  \label{eqn:shooting-a}\\
  A_0''(r)+\frac{2}{r}A_0'(r)+2ef(r)^2g(r)&=0,
  \label{eqn:shooting-b}
\end{align}
where we have defined $g(r)=\omega-eA_0(r)$.  There are several approaches to
finding solutions which satisfy the coupled equations
\eqref{eqn:shooting-a}--\eqref{eqn:shooting-b} such as shooting \cite{Lee1989},
relaxation \cite{Gulamov2015}, or via mapping from the profiles of non-gauged
\qballs{} \cite{Heeck2021}. Here we utilize an iterative shooting procedure to
numerically determine $f(r)$ and $A_0(r)$ which satisfy
\eqref{eqn:shooting-a}--\eqref{eqn:shooting-b} to a good approximation. Further
details about this technique are provided in \cite{Kinach2023}.

\section{\label{sec:numerics} Numerical Approach}

As a starting point for our evolution, we consider the line element 
\begin{equation}
  ds^2=-dt^2+d\rho^2+\rho^2d\varphi^2+dz^2
\end{equation}
where $(t,\rho,\varphi,z)$ are the standard cylindrical coordinates.  
Further,
we impose axisymmetry on the system by requiring all dynamical variables
to be $\varphi$-independent.
This is done purely to reduce the computational cost of
modelling the system in fully three spatial dimensions.  With this choice, the
equations of motion \eqref{eqn:eom-a}--\eqref{eqn:eom-b} can be expressed as a
set of six coupled non-linear partial differential equations; these equations
are identical to those listed in the appendix of our previous paper
\cite{Kinach2023}.  Working in the Lorenz gauge, the equations of motion are
supplemented with the gauge condition
\begin{equation}
  \label{eqn:lorenz}
  \nabla_\mu A^\mu=0,
\end{equation}
and the 
equations
\begin{align}
  \nabla_i E^i&=ej^0\, , \label{eqn:divE}\\
  \nabla_i B^i&=0\, \label{eqn:divB},
\end{align}
where $E^i$ and $B^i$ are the (three-dimensional) electric and magnetic field
vectors, respectively, whose components are determined via the electromagnetic
field tensor, $F_{\mu\nu}$.  Together, the equations
\eqref{eqn:lorenz}--\eqref{eqn:divB} act as additional constraints on the
evolution: it is expected that a numerical solution to the equations of motion
will approximately satisfy these constraint equations at any given time.

In order to construct initial data which is suitable for studying head-on
collisions, we interpolate a pair of spherically-symmetric gauged \qball{}
solutions in the $\rho-z$ plane using Neville's algorithm to fourth-order
in the mesh spacing
\cite{Press2007}. The center of each \qball{} is chosen to coincide with the
line $\rho=0$ in order to preserve the spherical symmetry of each \qball{} in
the binary.  Each \qball{} is also given an initial displacement along the
$z$-axis so that the binary is well-separated at the initial time. Finally, we
apply a Lorentz boost to each \qball{} along the $z$-direction at a
relativistic speed $v$ (where $v=1$ corresponds to the speed of light in our
units) so that they travel toward each other. After these operations, the field
variables $f\in \{\phi,\partial_t \phi,A_\mu,\partial_t A_\mu\}$ are
initialized according to the linear superposition
\begin{equation}
  \label{eqn:superposition}
  f(\rho,z)=f_A(\rho,z)+f_B(\rho,z),
\end{equation}
subject to the condition
\begin{equation}
  \label{eqn:overlap}
  f_A(\rho,z)\cdot f_B(\rho,z)\approx 0,
\end{equation}
where the subscripts $\{A,B\}$ identify each individual \qball{} in the
binary.

Practically speaking, the condition \eqref{eqn:overlap} is not trivial to
satisfy in general. While the scalar field falls off exponentially away from
the \qball{} center (thereby satisfying the condition even at modest separation
distances), the same cannot be said for the gauge field, which falls off like
$1/r$. This long-range behaviour inherently introduces violations of the
constraint equations \eqref{eqn:lorenz}--\eqref{eqn:divB} when the gauge fields
of each \qball{} significantly overlap. The magnitude of this violation depends
on several factors such as the initial separation distance, the boost velocity,
and the total charge of the constituent \qballs{}. To deal with this problem,
we implement an FAS multigrid algorithm \cite{Press2007} to re-solve the
equations \eqref{eqn:divE}--\eqref{eqn:divB} at the initial time and
minimize the constraint violation for arbitrary superpositions of the form
\eqref{eqn:superposition}. 
We also monitor the residuals of the constraint equations
\eqref{eqn:lorenz}--\eqref{eqn:divB} during the evolution to ensure that they
do not grow significantly over the timescales under consideration.

For the purposes of this work, we choose several representative examples of
gauged \qball{} solutions to act as initial data for the colliding binaries.
The properties of these solutions are listed in Table \ref{table:solns}. In our
simulations, we consider two different possibilities for the scalar field
potential $V(|\phi|)$ in the model \eqref{eqn:gauged-lagr}. These are
\begin{align}
  V_\text{log}(|\phi|)&=-\mu^2|\phi|^2\ln(\beta^2|\phi|^2),\label{eqn:log}\\
  V_\text{6}(|\phi|)&=m^2|\phi|^2-\frac{k}{2}|\phi|^4+\frac{h}{3}|\phi|^6,\label{eqn:poly}
\end{align}
where $\mu$, $\beta$, $m$, $k$, and $h$ are real, positive parameters. In Table
\ref{table:solns}, the solutions pertaining to the logarithmic potential
\eqref{eqn:log} are named LogA, LogB and LogC while the solutions due to the polynomial
potential \eqref{eqn:poly} are named PolyA and PolyB. These solutions, which
are known to be stable against axisymmetric perturbations \cite{Kinach2023}, are
specifically chosen to illustrate the range of dynamical features associated
with head-on collisions of gauged \qballs{}. We emphasize that aside from the
examples listed in Table \ref{table:solns}, we have also studied collisions
involving several other configurations and find the dynamics to be consistent
with the results reported below. 

In addition to varying the scalar potential, we also adjust the values of the
electromagnetic coupling constant $e$, the initial velocity $v$, the relative
phase difference $\alpha$, and the relative sign of the Noether charge $Q$ for
the colliding \qballs{}. The value of $\alpha$ is set through a simple
modification of the spherical \qball{} ansatz
\eqref{eqn:sph-ansatz-f}:
\begin{equation}
 \label{eqn:sph-ansatz-mod}
  \phi(t,\vec{x})=f(r)\,e^{\epsilon(i \omega t) + i\alpha},
\end{equation}
where $\alpha\in[0,\pi]$ and $\epsilon=\pm 1$. Since we only consider
collisions between \qballs{} with identical $\omega$, the value of $\alpha$
determines the relative difference in phase between the colliding \qballs{}
prior to the moment of impact. The sign of $\epsilon$, meanwhile, provides a
mechanism through which we can study both \qball{}/\qball{} and
\qball/anti-\qball{} collisions.  This can be understood from the fact that the
sign of the Noether charge $Q$ (and the sign of the electric charge $Q_e=eQ$)
of a gauged \qball{} is connected to the sign of the oscillation frequency
$\omega$ \cite{Gulamov2015}. Therefore, adjusting the sign of $\epsilon$ for
one \qball{} in the binary (as well as taking $A_0(r)\rightarrow-A_0(r)$ in
\eqref{eqn:sph-ansatz-A}) effectively flips the sign of its charge,
allowing us to superpose initial data of equal or opposite charge as desired.

After specifying the initial data at $t=0$, we proceed by evolving the system
forward in time. To facilitate this, we invoke a coordinate transformation
$x^\mu=(t,\rho,z)\rightarrow x^{\mu'}=(t,P,Z)$ according to
\begin{align}
  \rho &= d\exp(cP) - d\exp(-cP), \label{eqn:compactP} \\
  z &= d\exp(cZ) - d\exp(-cZ), \label{eqn:compactZ}
\end{align}
where $c$ and $d$ are positive, real parameters. With appropriate choice of $c$
and $d$, the transformation \eqref{eqn:compactP}--\eqref{eqn:compactZ} remains
approximately linear near the origin while becoming increasingly compactified
at large coordinate values.  This is an attractive feature for our numerical
domain because it allows us to resolve the dynamics at large length scales
without incurring an excessive computational cost.  To perform the evolution in
this coordinate system, we use a second-order Crank-Nicolson finite-difference
scheme implemented with fourth-order Kreiss-Oliger dissipation as a smoothing
operator. A modified Berger-Oliger adaptive mesh refinement (AMR) algorithm
\cite{Pretorius2006} is used to dynamically increase the numerical resolution
of our simulations in the regions of greatest interest.  For all results
presented below, the base grid is taken to be 129 by 257 grid points in $\{P,
Z\}$ with up to 8 levels of additional mesh refinement at a refinement ratio of
2:1. We choose a Courant factor of $\lambda = dt/ \min\{dP, dZ\} = 0.25$.  At
the outer boundaries, we impose outgoing (Sommerfeld) boundary conditions in
order to accommodate the long-range behaviour of the electromagnetic field and
reduce the effects of spurious boundary noise. In addition, we apply reflective
or anti-reflective boundary conditions as necessary along the axis of symmetry
in order to enforce regularity.

For numerical convenience, we choose $\mu=\beta=m=k=1$ and $h=0.2$ in
\eqref{eqn:log}--\eqref{eqn:poly} following our previous work
\cite{Kinach2023}.  We select $c=0.05$, $d=10$ in
\eqref{eqn:compactP}--\eqref{eqn:compactZ} and set the domain boundaries to
span at least $\{P\,:\,0\le P \le 50\}$ and $\{Z\,:\,-50\le Z \le 50\}$ which
corresponds to $\{\rho\,:\,0\le\rho\lesssim 121\}$ and $\{z\,:\,-121\lesssim z
\lesssim 121\}$ in the original coordinate system. With this choice, we find
the numerical domain to be large enough to capture the 
relevant post-collision dynamics of the \qballs{}.
We emphasize that while all evolutions have been performed
using the compactified coordinates $P$ and $Z$, we will hereafter present all
results using the linear coordinates $\rho$ and $z$.  This is done primarily to
facilitate the interpretation of the results. Finally, since the numerical code
is identical to the one used in \cite{Kinach2023} (aside from applying the
coordinate transformation \eqref{eqn:compactP}--\eqref{eqn:compactZ} and the
generation of binary initial data), we refer the reader to \cite{Kinach2023}
for issues of code validation such as convergence and independent residual
tests.

{ \setlength\extrarowheight{2pt} \begin{table*}
  \begin{tabular}{|c|c|c|c|c|c|c|c|c|} \hline
    Solution     & $e$     & $|\phi(0,0)|$      & $A_0(0,0)$            & $\omega$  & $E$               & $|Q|$            \\ \hline \hline
    LogA         & $0.1$   & $0.3669$           & $2.697\times10^{-2}$  & $2.003$   & $6.769$           & $3.006$          \\ 
    LogB         & $0.1$   & $1.627$            & $0.2682$              & $1.027$   & $45.45$           & $30.03$          \\ 
    LogC         & $1.1$   & $0.6461$           & $1.383$               & $2.522$   & $52.08$           & $22.37$          \\ 
    PolyA        & $0.02$  & $2.062$            & $0.4353$              & $0.6587$  & $476.4$           & $582.9$          \\ 
    PolyB        & $0.17$  & $1.973$            & $2.515$               & $0.9976$  & $405.1$           & $387.5$          \\ 
    \hline
  \end{tabular}
  \caption{Table of several gauged \qball{} solutions used in our collision
  simulations. The solutions LogA, LogB and LogC correspond to the logarithmic potential
  \eqref{eqn:log} while PolyA and PolyB correspond to the polynomial potential
  \eqref{eqn:poly}. From left to right, the remaining columns indicate the
  value of the electromagnetic coupling constant $e$, the initial central value
  of the scalar field $|\phi(0,0)|$, the initial central value of the gauge
  field $A_0(0,0)$, the \qball{} oscillation frequency $\omega$, the total
  energy $E$ of the solution (when stationary), and the total Noether charge
  $|Q|$ of the solution.}
\label{table:solns} \end{table*} }

\section{\label{sec:results} Numerical Results}

We now describe the results of our numerical experiments. In our collision
simulations, we consider the effects of the following parameters on the
resulting dynamics: gauge coupling strength $e$, collision velocity $v$,
relative phase difference $\alpha$, and relative sign of the Noether charge
$Q$.  In most cases, we restrict the collision velocity to the range
$0.1 \le v \le 0.9$ and the phase difference to
$\alpha\in\{0,\pi/4,\pi/2,3\pi/4,\pi\}$, though in some cases we explore beyond
these values to get a complete picture of the dynamics.  Further, we test the
effects of the choice of scalar potential (logarithmic \eqref{eqn:log} versus
polynomial \eqref{eqn:poly}) as well as the difference between colliding
\qballs{} of equal charge and opposite charge. We note that for all simulations
presented below, the constituent \qballs{} are always composed of identical
charge magnitudes (i.e., we do not present any results for collisions between
\qballs{} of differing $|Q|$). For comparison purposes, we first explore the
results at small gauge coupling.  We then move on to the case where the gauge
field is strongly coupled to highlight the salient dynamics.  For presentation
purposes, we have relegated some plots of the dynamics in this section to
App.~\ref{sec:suppfigs}.

We provide in Table \ref{table:summary} a broad, high-level overview of the
main results of our numerical experiments. We will devote the remainder of this
work to discussing the various phenomena which are reflected in the table.

{ \setlength\extrarowheight{2pt} 
\begin{table*}[]
\begin{tabular}{|ccc|cc|}
\hline
\multicolumn{3}{|c|}{Collision Parameters}                                                                                                              & \multicolumn{2}{c|}{Result}                                                                                                                       \\ \hline
\multicolumn{1}{|c|}{Relative Charge $Q$}        &
  \multicolumn{1}{c|}{Phase Difference $\alpha$}           & Collision
  Velocity $v$                    & \multicolumn{1}{c|}{Small $e$}
  & Large $e$
  \\ \hline \hline
\multicolumn{1}{|c|}{\multirow{11}{*}{Equal $Q$}} &
  \multicolumn{1}{c|}{\multirow{3}{*}{$\alpha=0$}}         & Low $v$
  & \multicolumn{1}{c|}{Merger}                           &
  \begin{tabular}[c]{@{}c@{}}Coulomb repulsion\\ (no collision)\end{tabular}          \\ \cline{3-5}
\multicolumn{1}{|c|}{}                           & \multicolumn{1}{c|}{}                                    & Intermediate $v$                          & \multicolumn{1}{c|}{Merger, fragmentation}            & Merger, fragmentation                                                                     \\ \cline{3-5}
\multicolumn{1}{|c|}{}                           & \multicolumn{1}{c|}{}                                    & High $v$                                  & \multicolumn{1}{c|}{Free-passage}                     & Fragmentation                                                                             \\ \cline{2-5}
\multicolumn{1}{|c|}{}                           &
  \multicolumn{1}{c|}{\multirow{3}{*}{$\alpha\in(0,\pi)$}} & Low $v$
  & \multicolumn{1}{c|}{\multirow{3}{*}{Charge transfer}} &
  \begin{tabular}[c]{@{}c@{}}Coulomb repulsion\\ (no collision)\end{tabular}          \\ \cline{3-3} \cline{5-5}
\multicolumn{1}{|c|}{}                           & \multicolumn{1}{c|}{}                                    & \multirow{2}{*}{Intermediate \& High $v$} & \multicolumn{1}{c|}{}                                 & \multirow{2}{*}{\begin{tabular}[c]{@{}c@{}}Charge transfer,\\ fragmentation\end{tabular}} \\
\multicolumn{1}{|c|}{}                           & \multicolumn{1}{c|}{}                                    &                                           & \multicolumn{1}{c|}{}                                 &                                                                                           \\ \cline{2-5}
\multicolumn{1}{|c|}{}                           &
  \multicolumn{1}{c|}{\multirow{3}{*}{$\alpha=\pi$}}       & Low $v$
  & \multicolumn{1}{c|}{\multirow{3}{*}{Phase repulsion}} &
  \begin{tabular}[c]{@{}c@{}}Coulomb repulsion\\ (no collision)\end{tabular}          \\ \cline{3-3} \cline{5-5}
\multicolumn{1}{|c|}{}                           & \multicolumn{1}{c|}{}                                    & \multirow{2}{*}{Intermediate \& High $v$} & \multicolumn{1}{c|}{}                                 & \multirow{2}{*}{Phase repulsion}                                                          \\
\multicolumn{1}{|c|}{}                           & \multicolumn{1}{c|}{}                                    &                                           & \multicolumn{1}{c|}{}                                 &                                                                                           \\ \hline
\multicolumn{1}{|c|}{Opposite $Q$}               & \multicolumn{1}{c|}{All $\alpha$}                        & All $v$                                   & \multicolumn{1}{c|}{Partial annihilation}             & \begin{tabular}[c]{@{}c@{}}Partial annihilation,\\ radiation emission\end{tabular}        \\ \hline
\end{tabular}
  \caption{Summary of the main dynamical results from our collision
  simulations. Shown are the observed collision outcomes (classified by either
  ``small" or ``large" values of the gauge coupling constant $e$) as a function
  of various collision parameters: the relative Noether charge $Q$ of the
  colliding binary (either equal or opposite), the relative phase difference
  $\alpha$, and the collision velocity $v$ (heuristically divided into
  ``low-velocity", ``intermediate-velocity", and ``high-velocity" regimes). We
  comment that the results listed in this table together capture the dynamics
  in both the logarithmic \eqref{eqn:log} and polynomial \eqref{eqn:poly}
  scalar field models. These results are explained in further detail throughout
  Sec.~\ref{sec:results}.}
\label{table:summary} \end{table*} }

\subsection{Small Gauge Coupling}

Here we consider collisions involving solutions LogA, LogB, and PolyA from
Table \ref{table:solns}. Since the strength of the gauge coupling is small in
these cases (see \cite{Gulamov2014} where this notion is made precise), it is
expected that the dynamics of gauged \qballs{} in this regime will be similar
to the dynamics of ordinary (non-gauged) \qballs{}.

Let us begin by discussing the effect of \qball{} velocity on the outcome of
the collision. In previous studies \cite{Axenides2000,Battye2000,Gutierrez2013}
it has been shown that the dynamics of equal-charge, non-gauged \qball{}
collisions can generally be divided into three regimes: (i) at low velocities,
a ``merger" regime wherein the \qballs{} tend to coalesce, (ii) at intermediate
velocities, a ``fragmentation" regime wherein the \qballs{} tend to break up
into smaller components, and (iii) at high velocities, an ``elastic" regime
wherein the \qballs{} tend to pass through each other virtually unscathed.  We
find that gauged \qball{} collisions with small gauge coupling are generally
consistent with these previous findings.

\begin{figure} [t]
\includegraphics[width=\columnwidth]{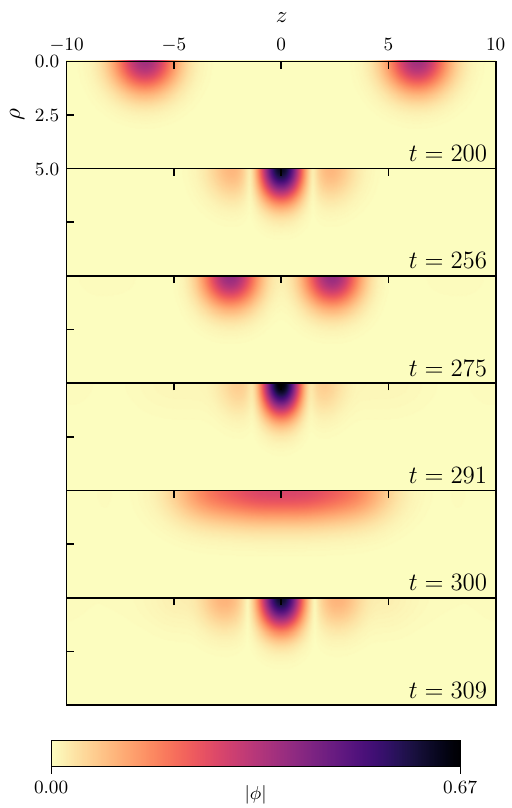}
  \caption{\label{fig:nongauged-merger}
  Evolution of the scalar field modulus $|\phi|$ for a collision of solutions
  of type LogA with equal charge, velocity $v=0.1$, and phase difference
  $\alpha=0$.  The \qballs{} collide at $t\approx250$ and repeatedly
  merge and separate. By $t\approx600$ (beyond what is shown here), the
  field configuration settles down into a single larger \qball{} which remains
  perturbed at the origin.}
\end{figure}

First, consider the low-velocity regime. In Fig.~\ref{fig:nongauged-merger}, we
plot the collision of two \qballs{} of type LogA (see Table \ref{table:solns})
with equal charge, velocity $v=0.1$, and phase difference $\alpha=0$. As the
\qballs{} collide, they merge temporarily before separating again and
propagating a short distance along the axis of symmetry. However, they have
insufficient kinetic energy to completely escape their mutual influence and
instead repeatedly merge and partially separate. Small amounts of scalar matter
are also released during this process. As the evolution proceeds, the field
configuration settles down into a single coherent merged state.  The final
\qball{} is of a larger total size than LogA and remains at the origin lightly
perturbed.

When boosted to velocities above a certain threshold, the colliding \qballs{}
have sufficient kinetic energy to avoid a merged final state (for LogA, the
velocity threshold is $v\gtrsim 0.125$). At these ``intermediate" velocities, a
significant quantity of the initial charge of each \qball{} continues
propagating along the axis of symmetry after the collision. These resulting
\qballs{} are highly perturbed and oscillatory. In most cases, this process
also results in some relic amount of charge left behind: the \qballs{} have
partially fragmented into smaller structures. These smaller \qballs{} may
either remain stationary at the origin or continue to propagate along the axis
of symmetry, lagging the main \qballs{} at a lower velocity. 
An example of such a collision for solution LogA at velocity $v=0.5$ is given
in App.~\ref{sec:suppfigs} (Fig.~\ref{fig:nongauged-fragment}).

\begin{figure} [h]
\includegraphics[width=\columnwidth]{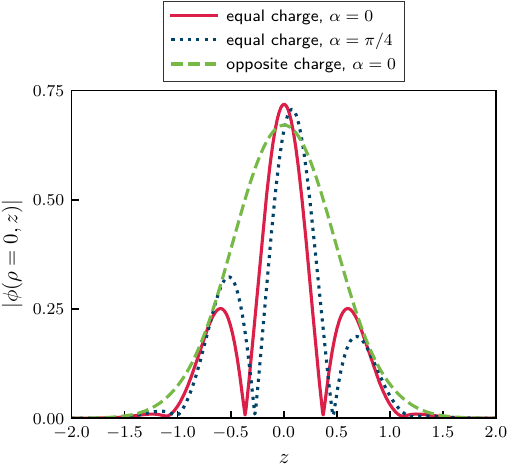}
  \caption{\label{fig:nongauged-interference}
  Profiles of the scalar field modulus $|\phi|$ evaluated along the axis of
  symmetry during collisions involving solution LogA with $v=0.9$.  Three cases
  are shown: an equal-charge collision with no phase difference ($\alpha=0$),
  an equal-charge collision with phase difference $\alpha=\pi/4$, and an
  opposite-charge collision with no phase difference ($\alpha=0$). In each
  case, the profile is shown at the moment $|\phi|$ reaches its maximal
  value. For collisions with equal charge, a destructive interference pattern
  forms at the moment of impact. For collisions with opposite charge, the
  interference pattern is purely constructive.}
\end{figure}

At the highest velocities, collisions between the \qballs{} are primarily
elastic and they emerge from the collision relatively unscathed.
An illustration of this phenomenon is given in App.~\ref{sec:suppfigs}
(Fig.~\ref{fig:nongauged-passthru}) for solution LogA at velocity $v=0.9$.
It is also in this regime that the wave-like nature of \qballs{} becomes
readily apparent through the appearance of interference fringes at the moment
of impact.  Plotted in Fig.~\ref{fig:nongauged-interference} are the
interference fringes observed for collisions of solution LogA at $v=0.9$. For
equal-charge collisions, a clear fringe pattern emerges with fringe spacing
inversely proportional to the collision velocity. Also shown are the effects of
opposite-charge and phase-difference collisions on the fringe pattern (to be
discussed below).

We now comment on the effects of phase difference on the collision dynamics.
Recall that a phase difference is introduced into the system by choosing
$\alpha\neq 0$ in \eqref{eqn:sph-ansatz-mod}. Since the colliding \qballs{} in
our study always have identical values of $\omega$, this phase difference is
preserved until the moment of impact regardless of the initial separation
distance or initial velocity. As reported previously \cite{Battye2000}, the
main effect of this phase difference is to induce charge transfer between the
colliding \qballs{}.  This behaviour can be understood in terms of relative
phase accelerations \cite{Battye2000} or the induced rate of change of momentum
for the colliding \qballs{} \cite{Bowcock2009}. Testing the effects of phase
difference
at $\alpha\in\{0,\pi/4,\pi/2,3\pi/4,\pi\}$, we find that charge transfer is
generally maximized at the lowest collision velocities and for small phase
differences, in agreement with previous studies.

\begin{figure} [t]
\includegraphics[width=\columnwidth]{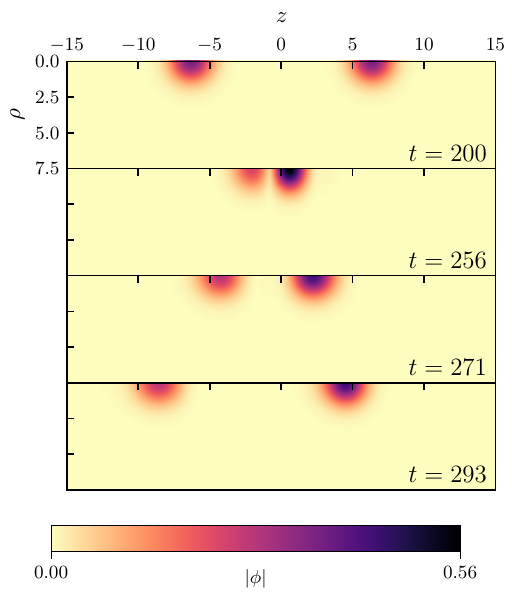}
  \caption{\label{fig:nongauged-transfer}
  Evolution of the scalar field modulus $|\phi|$ for a collision of solutions
  of type LogA with equal charge, velocity $v=0.1$, and phase difference
  $\alpha=\pi/4$. After colliding at $t\approx250$, the
  \qball{} with leading phase (left) transfers charge to the \qball{} with
  lagging phase (right).  After the collision, the \qballs{} have disparate
  velocities.}
\end{figure}

Plotted in Fig.~\ref{fig:nongauged-transfer} is the collision of solution LogA
at a velocity of $v=0.1$ and a phase difference $\alpha=\pi/4$. Initially, the
\qballs{} are of equal charge.  At the moment of impact, the \qball{} with
lagging phase (rightmost \qball{} in the figure) suddenly gains charge from the
\qball{} with leading phase (leftmost \qball{}). Since \qballs{} are extended
structures, it can be difficult to precisely determine the total charge $Q$
contained in the resulting objects. However, by integrating $Q$ in the
half-volumes $z>0$ and $z<0$ after the collision takes place, we can estimate
by the deviation from symmetry that approximately 18\% of the charge is
transferred during this process. We note that the total charge $Q$ over the
simulation domain remains conserved to within 0.1\% during the evolution. In
addition to charge transfer, we observe that the velocities of the resultant
\qballs{} after the collision are no longer identical: the smaller \qball{}
moves faster than the larger one. This can be understood as a straightforward
consequence of linear momentum conservation.

At intermediate velocities, we observe the same qualitative behaviour, though
with the amount of charge transfer reduced (for instance, only ${\sim}\,7\%$ is
transferred at $v=0.5$, $\alpha=\pi/4$ for solution LogA). In some cases,
the charge transfer at these velocities is accompanied by the formation of one
or more smaller \qballs{} which remain along the axis of symmetry after the
collision and lag the main \qballs{}, being slightly perturbed. At the highest
velocities, the charge transfer is minimal (for instance, ${\sim}\,1\%$
or less of the charge is transferred with $v\gtrsim0.9$, $\alpha=\pi/4$ for solution
LogA) and no significant smaller \qballs{} are formed during the collision.
However, the phase difference still manifests through a distortion of the
interference fringes as illustrated in Fig.~\ref{fig:nongauged-interference}.

A notable exception to the charge transfer phenomenon occurs for completely
out-of-phase collisions ($\alpha=\pi$). In this case, the \qballs{} exhibit a
purely repulsive interaction as they ``bounce" off each other.  At the moment
of impact, the \qballs{} are compressed in the boost direction and the value of
$|\phi|$ temporarily grows by an amount which is proportional to the collision
velocity. There is no charge transfer observed: the half-volumes $z>0$ and
$z<0$ contain an identical amount of charge for all time. Note that this
repulsive behaviour for out-of-phase collisions has also been observed in other
soliton models \cite{Schwabe2016, Cardoso2016}.

\begin{figure}
\includegraphics[width=\columnwidth]{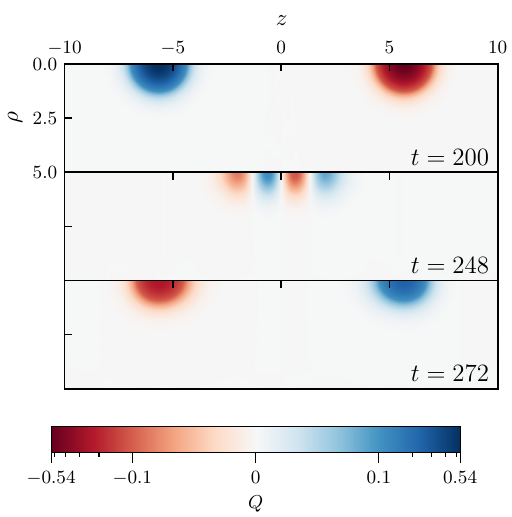}
  \caption{\label{fig:nongauged-annihilation}
  Evolution of the Noether charge $Q$ for a collision of solutions of type LogA
  with opposite charge, velocity $v=0.1$, and phase difference $\alpha=0$. The
  \qballs{} collide at $t\approx248$ and partially annihilate charge. After the
  collision, the resultant \qballs{} pass through each other and continue
  propagating along the axis of symmetry with a larger velocity.
  Note that a hybrid colormap is used: charge values below $|Q|=0.1$ are mapped
  linearly to zero while values above this threshold are mapped
  logarithmically to the charge maximum.}
\end{figure}

We now discuss collisions of oppositely-charged \qballs{}. These are the
ones for which $\epsilon=-1$ in equation \eqref{eqn:sph-ansatz-mod} for
one of the \qballs{} in the binary, resulting in a system composed of a
gauged \qball{} and gauged anti-\qball{}. These collisions are
predominantly characterized by the possibility of charge annihilation at
the moment of impact, with the amount of annihilation depending on the
collision velocity. For example, an opposite-charge collision
corresponding to solution LogA at $v=0.1$ results in ${\sim}\,48\%$
of the charge annihilated. This situation is depicted in
Fig.~\ref{fig:nongauged-annihilation}. The remaining charge emerges from
the collision in the form of smaller \qballs{} with a larger velocity.
In addition, the relatively violent dynamics that occur during the
annihilation leave them highly perturbed and oscillatory after the
collision.  

Charge annihilation during opposite-charge \qball{} collisions is also observed
at larger velocities, though the amount of annihilation is reduced. For example,
the amount of charge annihilated is ${\sim}\,15\%$ at $v=0.3$ and ${\sim}\,7\%$
at $v=0.5$ for solution LogA. In addition, the collision at these
larger velocities is sometimes accompanied by the creation of smaller \qballs{}
remnants which remain along the axis of symmetry.  At the highest velocities,
the \qball{}/anti-\qball{} interaction results in very little annihilation (for
example, only ${\sim}\,1\%$ of charge is annihilated at $v=0.9$). There
are also fewer \qball{} remnants produced along the axis of symmetry and the
fields interfere constructively at the moment of impact (see
Fig.~\ref{fig:nongauged-interference}).

We have also tested the effects of phase difference on \qball{}/anti-\qball{}
collisions, finding that it has a minimal influence on the dynamics. Charge
transfer is not observed and the amount of annihilation is not significantly
altered compared to the $\alpha=0$ case.

Thus far, we have only discussed the dynamics associated with solution
LogA. Now we turn to solution LogB in Table \ref{table:solns}. In this case, we
find that a generic outcome of the collision is that the field values tend to
grow without bound until the evolution becomes singular. This occurs even when
the calculation is repeated using additional levels of mesh refinement.
As discussed in \cite{Kinach2023}, we can understand this behaviour as a
consequence of the logarithmic potential \eqref{eqn:log} being unbounded from
below. In particular, for large scalar field values (such as those achieved at
the moment of impact), the potential term $V(|\phi|)$ in \eqref{eqn:emt} can
become negative and may dominate over the other energies in the system. This
can lead to the energy density becoming locally negative in the region of large
$|\phi|$. At the same time, the energy density in other areas of the domain
must grow so that the total integrated energy remains conserved to a positive
quantity. This reciprocal process can result in runaway field growth which
quickly causes the evolution to become singular.  Due to such pathological
effects, we do not consider collisions of \qballs{} with sizes much larger than
that of LogA for $e=0.1$ in the logarithmic model.

To conclude this section, let us consider the collision dynamics under the
polynomial potential \eqref{eqn:poly}. For this purpose, we will use solution
PolyA in Table \ref{table:solns} as an illustrative example. Much like what is
observed for solution LogA, we find that equal-charge collisions at low
velocities are characterized by a merger regime. Notably, the range of
velocities for which the \qballs{} merge is quite large---in our experiments,
merging occurs for $v\lesssim0.7$. At higher collision velocities, the \qballs{}
have sufficient kinetic energy to escape the merged state and continue
propagating along the axis of symmetry after passing through each other.  This
is accompanied by a small portion of field content radiating away from the
\qballs{} after the moment of impact.
We have also tested the effects of phase-difference and opposite-charge
collisions involving solution PolyA, finding evidence for charge transfer and
annihilation similar to what has been previously discussed.

\subsection{Large Gauge Coupling}

We now turn to collisions involving solutions LogC and PolyB from Table
\ref{table:solns}. Unlike the collisions discussed in the previous section,
these solutions involve a gauge coupling which is comparable in magnitude to
the scalar potential parameters.  We therefore expect that electromagnetic
effects may have a non-trivial impact on the dynamics.

Once again, we begin by discussing the effect of the initial velocity on the
outcome of the collision. Since the \qballs{} can now carry a significant
amount of electric charge, the long-range Coulomb force can influence the
dynamics prior to the moment of impact. If the colliding \qballs{} have equal
charge, this results in deceleration and a corresponding decrease in their
effective velocity before impact. If the colliding \qballs{} have opposite
charge, the result is acceleration which increases the effective velocity. In
order to fully capture this behaviour, it would be preferable to initialize the
boosted \qballs{} at $z=\pm \infty$ and let them travel toward each other.
However, limitations in computational resources make it unfeasible to
initialize the fields at arbitrarily large separation distances, so instead we
initialize the \qballs{} at $z=\pm25$ for a given boost. As mentioned
previously, we use a multigrid solver to remedy the unphysical constraint
violations which may result from 
a simple superposition of the scalar and
electromagnetic fields.
In what follows, we will refer to the collision velocity as the velocity at
which the \qballs{} are initialized at $z=\pm25$ rather than their effective
velocity at the moment of impact.

\begin{figure}
\includegraphics[width=\columnwidth]{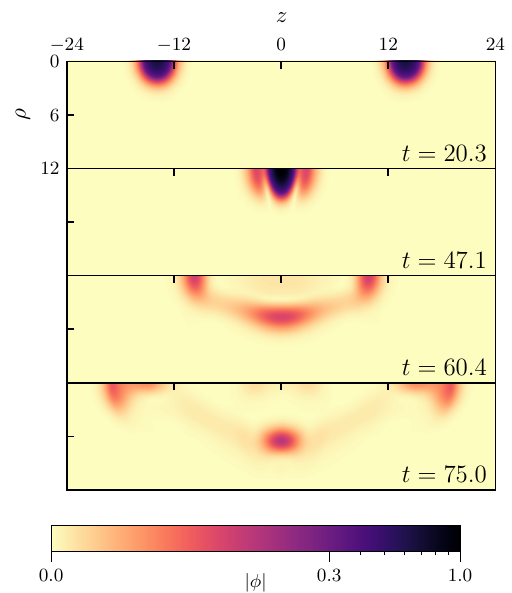}
  \caption{\label{fig:nongauged-rightangle}
  Evolution of the scalar field modulus $|\phi|$ for a collision of solutions
  of type LogC with equal charge, velocity $v=0.55$, and phase difference
  $\alpha=0$. The \qballs{} collide at $t\approx45$. After the collision, the
  field content contains a mixture of on-axis and off-axis components.  Note
  that a hybrid colormap is used: field values below $|\phi|=0.3$ are mapped
  linearly to zero while values above this threshold are mapped logarithmically
  to the field maximum.}
\end{figure}

To proceed with the analysis, we consider the solution LogC in Table
\ref{table:solns}.
Unlike what has been discussed in the case of LogA (corresponding to small
gauge coupling), the dynamics of solution LogC during equal-charge collisions
cannot be cleanly divided into a merger, fragmentation, and elastic regime. At
low velocities, we find instead that the Coulomb repulsion is strong enough to
completely prevent the scalar fields of each \qball{} from significantly
interacting. This causes the \qballs{} to decelerate as they approach each other,
reach a turning point of vanishing velocity, and then accelerate away in the
opposite direction. This behaviour is found to occur for $0<v\lesssim0.3$.
At velocities $v\gtrsim0.3$, the \qballs{} have sufficient kinetic energy
to overcome the Coulomb repulsion and will eventually collide. In these
situations, the general outcome is fragmentation of the gauged \qball{} into
smaller components. Plotted in Fig.~\ref{fig:nongauged-rightangle}
is the collision of solution LogC at $v=0.55$. In contrast to the case of
small gauge coupling (where no off-axis remnants were observed in the
logarithmic model), here we see the formation of a distinct off-axis component
which propagates outward before collapsing back onto the axis of symmetry at
late times. As noted in \cite{Kinach2023}, these off-axis components represent
ring-like structures in three-dimensions which we call ``gauged \qring{}s". In
addition to the ring, a significant portion of the field content also passes
through the origin and continues propagating along the axis of symmetry while
being highly perturbed.

\begin{figure} [t]
\includegraphics[width=\columnwidth]{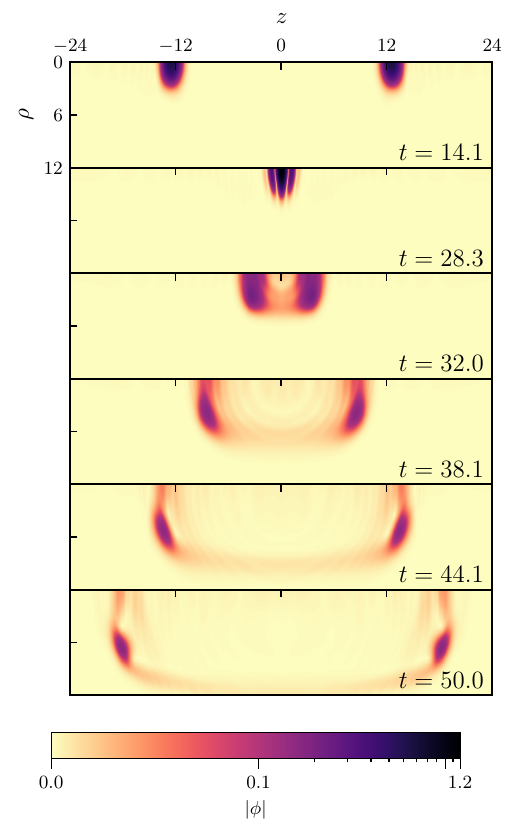}
  \caption{\label{fig:gauged-Qrings}
  Evolution of the scalar field modulus $|\phi|$ for a collision of
  solutions of type LogC with equal charge, velocity $v=0.9$, and phase
  difference $\alpha=0$.  The \qballs{} collide at $t\approx27$. After
  the collision, a scalar radiation pattern appears (fourth panel) and
  the field content predominantly takes the form of two \qrings{}.  Note
  that a hybrid colormap is used: field values below $|\phi|=0.1$
  are mapped linearly to zero while values above this threshold are
  mapped logarithmically to the field maximum.}
\end{figure}

At the highest velocities, the colliding \qballs{} form a clear destructive
interference pattern analogous to that seen for the case of small gauge
coupling (Fig.~\ref{fig:nongauged-interference}). However, after the collision,
the fields emerge primarily in the form of \qrings{} which propagate away from
the axis of symmetry. In addition, a scalar radiation pattern can be observed
in the vicinity of the origin. This situation is depicted in
Fig.~\ref{fig:gauged-Qrings} for solution LogC at $v=0.9$, and 
we have found this phenomenon to be present
up to a collision velocity of at least $v=0.95$. 
This 
contrasts what is observed for non-gauged \qballs{} where high-velocity
collisions primarily exhibit free-passage behaviour. 
Although computational
constraints prevent us from exploring boosts much beyond this range (in part
due to the extreme field gradients of the boosted \qballs{} at these velocities),
one can conclude that high-velocity collisions of gauged \qballs{} can be
considerably less elastic than collisions of their non-gauged counterparts.

Another challenge is to determine the ultimate fate of the observed \qrings{}.
While we have made some effort to track the long-term evolution of these
structures, the nature of the collision tends to see these remnants propagating
away at large velocities and reaching large coordinate distances. 
While the change
of coordinates \eqref{eqn:compactP}--\eqref{eqn:compactZ} can prevent these
components from exiting the domain entirely, they become increasingly
compactified as the evolution proceeds. When combined with our use of
Kreiss-Oliger dissipation for numerical stability, this effectively decreases
the numerical resolution of our simulations and increases the global error (as
measured, for instance, by an increase in the total constraint violation). As
such, it is difficult to conclusively determine the long-term behaviour of
these structures far from the origin, but we make the general observation that
they tend to reach a maximum radius before collapsing back inward toward the
axis of symmetry. We therefore conjecture that the gauged \qrings{} formed in
this way are transient objects (even if the growth of error prevents us from
making this statement definitively).

Next, we discuss the effects of phase difference for collisions involving
solution LogC. Similar to the case of non-gauged \qballs{}, the main effect of
altering the phase is to induce charge transfer during the collision.  However,
the large electric charge associated with LogC produces several novel effects.
The first is the absence of charge transfer at small collision velocities
$v\lesssim0.3$.  Similar to the case when $\alpha=0$, the Coulomb repulsion
prevents the scalar field of each \qball{} from significantly interacting and
so the charge transfer process is never observed. At larger velocities, the
\qballs{} have sufficient kinetic energy to fully interact and the result is a
net transfer of charge in a manner similar to the case of small gauge coupling.

\begin{figure} [t]
\includegraphics[width=\columnwidth]{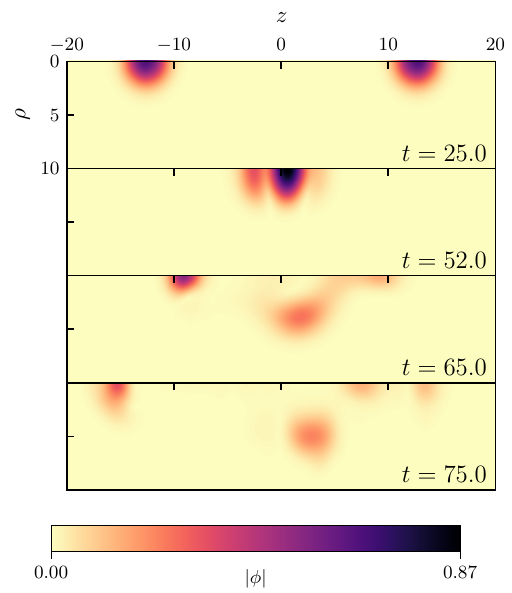}
  \caption{\label{fig:gauged-transfer}
  Evolution of the scalar field modulus $|\phi|$ for a collision of
  solutions of type LogC with equal charge, velocity $v=0.5$, and phase
  difference $\alpha=\pi/4$. The \qballs{} collide at $t\approx50$ and
  transfer charge (as can be seen in the second panel). After the
  collision, the larger \qball{} created in this process quickly breaks
  apart into smaller components which propagate on and away from the
  axis of symmetry. The smaller \qball{} travels toward $z=-\infty$ while
  highly perturbed.}
\end{figure}

One significant difference between charge transfer in the small- and
large-coupling case is the final fate of the \qballs{} after the collision. In
the case of small gauge coupling, the \qballs{} typically propagate away after
the collision and retain a coherent shape (though occasionally leaving behind a
small remnant \qball{} along the axis of symmetry).  However, for the case of
solution LogC (for example), the most common outcome is that the \qballs{}
created during the charge transfer process will quickly break apart into
smaller components.  This phenomenon is depicted in
Fig.~\ref{fig:gauged-transfer} for a collision involving solution LogC with a
phase difference of $\alpha=\pi/4$ and velocity $v=0.5$. Initially, the
\qballs{} are Lorentz-boosted toward each other and collide at $t\approx50$. In
this process, approximately 35\% of the charge is transferred. As the larger
\qball{} is formed, it is also highly perturbed, inducing its decay into
smaller \qballs{} and \qrings{}.  Depending on the collision parameters, this
instability can manifest in a number of different ways such as by breaking
apart into smaller \qballs{}, into \qrings{}, or into a combination of
\qballs{} and \qrings{}.  This phenomenon is presumably due to the reduced
parameter space of stable solutions which are allowed when the gauge coupling
is large \cite{Kinach2023}.

In general, we find that the charge transfer is maximal at intermediate
velocities $0.4\lesssim v\lesssim 0.6$ for solution LogC. At higher
velocities, the effect is still observed but the amount of charge
transfer is reduced (for example, the collision of solution LogC at
$v=0.7$, $\alpha=\pi/4$ results in ${\sim}\,10\%$ of the charge
transferred  while the same collision at $v=0.9$ results in only
${\sim}\,1\%$ transferred).  At these higher velocities, the charge
transfer manifests through slight asymmetries in the size and trajectory
of the \qring{} pattern. An example of this behaviour for solution LogC
at $v=0.7$, $\alpha=\pi/4$ is given in App.~\ref{sec:suppfigs}
(Fig.~\ref{fig:gauged-Qrings2}).

We have tested the amount of charge transfer at different phase differences in
the range $\alpha\in(0,\pi)$, finding that the transfer is maximal for $\alpha
\lesssim \pi/4$. The general phenomena associated with charge transfer is
similar for all $\alpha$ tested, though the individual dynamics may differ
slightly depending on the collision parameters.  However, one exception to the
previously-described behaviour is for the case of $\alpha=\pi$. Similar to what
has been observed for small gauge coupling, these out-of-phase \qballs{} tend
to experience a total repulsion at the moment of impact: the fields are
momentarily compressed before the \qballs{} ``bounce back" and form \qballs{}
or \qrings{} in manner symmetric about $z=0$ (i.e., there is no charge
transfer).

Finally, let us discuss \qball{}/anti-\qball{} interactions at large gauge
coupling. As was the case for small gauge coupling, the general outcome of such
collisions is the annihilation of charge.  However, unlike the case for
equal-charge collisions, the oppositely-charged \qballs{} now experience an
attractive Coulomb force which leads to acceleration prior to the moment of
impact; this effect is most noticeable at low velocities. This can lead to an
increase in the effective collision velocity as discussed previously.

\begin{figure}
\includegraphics[width=\columnwidth]{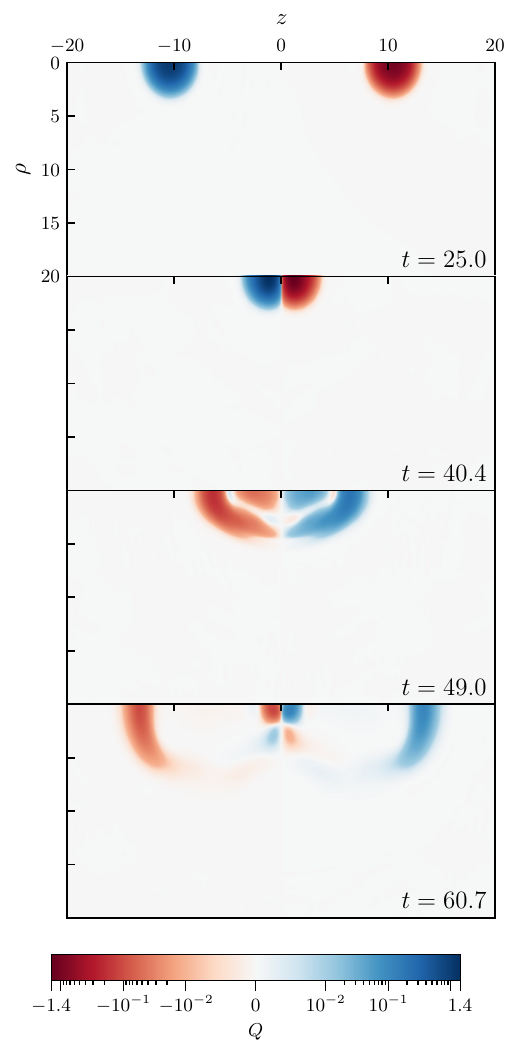}
  \caption{\label{fig:gauged-EMpulse2}
  Evolution of the Noether charge $Q$ for a collision of solutions of type LogC
  with opposite charge, velocity $v=0.6$, and phase difference $\alpha=0$. The
  \qballs{} collide at $t\approx40$ and partially annihilate charge. After the
  collision, a significant portion of the charge content continues propagating
  along the axis of symmetry while a remnant of mixed positive and negative
  charge is left behind at the origin. Note that a hybrid colormap is used:
  charge values below $|Q|=10^{-2}$ are mapped linearly to zero while values
  above this threshold are mapped logarithmically to the charge maximum.}
\end{figure}

\begin{figure}
\includegraphics[width=\columnwidth]{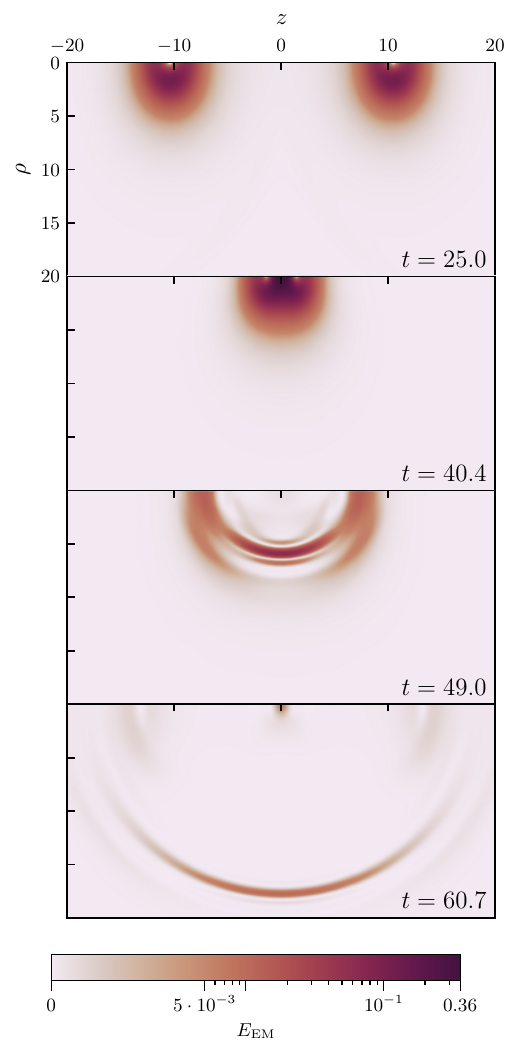}
  \caption{\label{fig:gauged-EMpulse}
  Evolution of the electromagnetic field energy $E_{\rm EM}$ for a collision of
  solutions of type LogC with opposite charge, velocity $v=0.6$, and phase
  difference $\alpha=0$. The \qballs{} collide at $t\approx40$ and partially
  annihilate charge. After the collision, a quasispherical pulse of electromagnetic
  energy emanates from the origin.  Note that a hybrid colormap is used: energy
  values below $E_{\rm EM}=5\cdot10^{-3}$ are mapped linearly to zero while values above
  this threshold are mapped logarithmically to the energy maximum.}
\end{figure}

Plotted in Fig.~\ref{fig:gauged-EMpulse2} is the Noether charge $Q$ for a
collision involving solution LogC with opposite charge, velocity $v=0.6$, and
phase difference $\alpha=0$. The \qballs{} collide at $t\approx 40$ and
partially annihilate. After the collision, a portion of each original \qball{}
continues propagating along the axis of symmetry. Additionally, there is a
small remnant of mixed charge left behind at the origin which
resembles in some ways a charge-swapping \qball{}
\cite{Copeland2014,Xie2021,Hou2022}. In this case, approximately ${\sim}\,53\%$
of the initial charge is annihilated during the collision.

The partial charge annihilation which occurs during a
\qball{}/anti-\qball{} collision can also result in the production of
electromagnetic radiation.  To observe this, we compute from
\eqref{eqn:emt} the energy contained in the electromagnetic field, which
can be written as
\begin{equation}
  E_{\rm EM}=\frac{1}{2}\left(|\vec{E}|^2+|\vec{B}|^2\right),
\end{equation}
where $\vec{E}$ and $\vec{B}$ are constructed from the components of the gauge
field $A_\mu$. 
The electromagnetic field energy for a collision involving solution LogC with
opposite charge, velocity $v=0.6$, and phase difference $\alpha=0$ (i.e., the
same collision as is plotted in Fig.~\ref{fig:gauged-EMpulse2}) is plotted in
Fig.~\ref{fig:gauged-EMpulse}. Initially, the motion of the charged \qballs{}
dominates the electromagnetic field energy. At the moment of impact, the
\qballs{} partially annihilate, converting a fraction of their total energy
into a pulse of electromagnetic energy which propagates away from the origin.
By comparing Fig.~\ref{fig:gauged-EMpulse2} and
Fig.~\ref{fig:gauged-EMpulse}, one can see that the outgoing pulse does not
correspond to any significant amount of charge. This fact supports our
interpretation of the pulse as representing electromagnetic radiation. We note
that we have not made an attempt to precisely quantify the amount of
electromagnetic radiation produced in this manner. This is due primarily to the
technical challenges associated with integrating the energy over arbitrary
subregions of the computational domain during adaptive, highly-parallelized
simulations.  However, we comment that the size of the electromagnetic pulse is
generally proportional to the amount of annihilation that occurs.
For illustrative purposes, we also plot in App.~\ref{sec:suppfigs}
(Fig.~\ref{fig:gauged-EMfield}) a representation of the electric and
magnetic fields for the collision depicted in
Fig.~\ref{fig:gauged-EMpulse2}/\ref{fig:gauged-EMpulse}.

\begin{figure}[t] 
\includegraphics[width=\columnwidth]{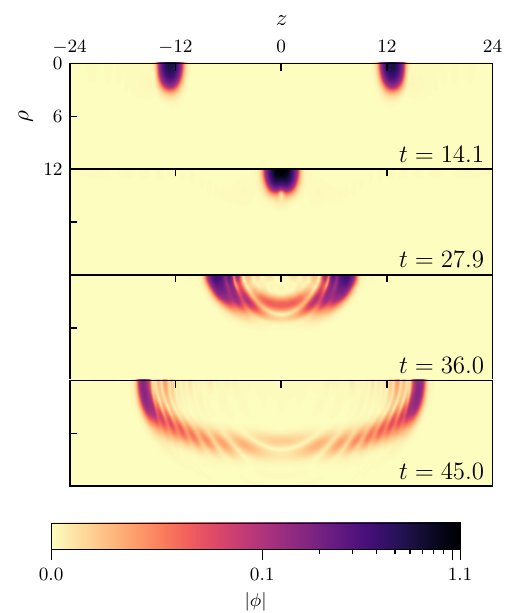}
  \caption{\label{fig:gauged-tail}
  Evolution of the scalar field modulus $|\phi|$ for a collision of
  solutions of type LogC with opposite charge, velocity $v=0.9$, and
  phase difference $\alpha=0$.  The \qballs{} collide at $t\approx27$
  and interfere constructively.  After the collision,
  the \qballs{} continue propagating along the axis of symmetry and
  carry a long ``tail" of scalar matter which exhibits an interference
  fringe pattern. Note that a hybrid colormap is used: field values
  below $|\phi|=0.1$ are mapped linearly to zero while values above this
  threshold are mapped logarithmically to the field maximum.}
\end{figure}

In the general case, we find that the dynamics of \qball{}/anti-\qball{}
interactions depend primarily on the collision velocity.
At the lowest velocities, the \qballs{} tend to pass through each other
after partially annihilating, then continue to travel along the axis of
symmetry while oscillating weakly. This process is often accompanied by
the partial fragmentation of the \qballs{} into a small number of
\qballs{} or \qrings{}. At intermediate velocities (e.g., $0.5\lesssim v
\lesssim 0.7$ for solution LogC), the collision becomes more violent:
the resulting \qballs{} and \qrings{} may be greater in number and more
strongly oscillatory after the collision.  It is also within this
intermediate regime that the charge annihilation is found to be maximal.
At the highest velocities (e.g., $v\gtrsim 0.7$ for solution LogC), the
outcome of the collision is once again dominated by two main \qballs{}
which continue propagating along the axis of symmetry.
These \qballs{} are accompanied by long ``tails" of the scalar field
which show a clear interference fringe pattern. This behaviour is
shown in Fig.~\ref{fig:gauged-tail} for solution LogC at $v=0.9$ with
opposite charges and $\alpha=0$. The amount of charge annihilation is
also reduced at high velocities (for example, only ${\sim}\,14\%$ of the
charge is annihilated for the collision depicted in
Fig.~\ref{fig:gauged-tail}).

We have also studied \qball{}/anti-\qball{} collisions of solution LogC
at various phase differences up to $\alpha=\pi$. We find that
the phase difference has a minimal effect and the phenomena associated with these
collisions resembles closely the $\alpha=0$ case. This suggests that the
collision dynamics of gauged \qballs{} with gauged anti-\qballs{} are
determined primarily by the collision velocity, in agreement with the
case of small gauge coupling.  It is interesting to note that we have
not observed any cases of total annihilation where the initial
\qballs{} are converted completely into radiation. Such a phenomena has
been observed in previous studies of non-gauged \qball{} collisions for
a small range of collision parameters \cite{Battye2000}. While total
annihilation may still be possible for the gauged case, our analysis
suggests that it might likewise occur for only a narrow range of
parameters.

\begin{figure}[t] 
\includegraphics[width=\columnwidth]{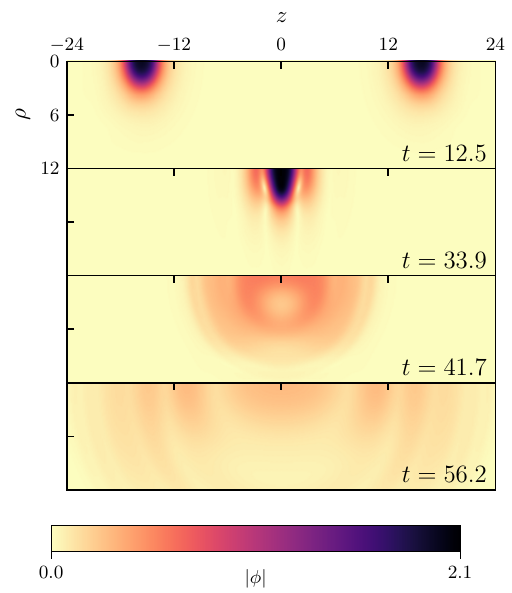}
  \caption{\label{fig:gauged-dissipate}
  Evolution of the scalar field modulus $|\phi|$ for a collision of
  solutions of type PolyB with equal charge, velocity $v=0.75$, and
  phase difference $\alpha=0$.  The \qballs{} collide at $t\approx33$
  and form a destructive interference pattern.  After the collision, it
  becomes difficult to distinguish any component of the field which
  clearly resembles a \qball{}. Instead, the field content appears to
  dissipate in the form of near-spherical waves which emanate from the
  origin.}
\end{figure}

We conclude this section by returning to collisions under the polynomial model
\eqref{eqn:poly}. For this purpose, we focus on solution PolyB in Table
\ref{table:solns}. This solution is notable in that it corresponds to a value
of the gauge coupling $e$ which is near the maximum allowed for the polynomial
potential, $e_{max}\approx0.182$ \cite{Loginov2020}.  Considering first the
equal-charge collisions of solution PolyB, we find once again that the
\qballs{} tend to repel at low velocities. This is in agreement with what has
been discussed previously for the logarithmic model.  However, for intermediate
velocities (e.g., $0.35\lesssim v\lesssim 0.6$), we observe that the colliding
\qballs{} can merge into a single \qball{} which remains at the origin. This is
accompanied by the emission of charge as the merged
\qball{} settles down into a near-stationary configuration.
At slightly higher velocities (e.g.  $0.65\lesssim v \lesssim 0.85$), the
\qballs{} do not form a single stable \qball{}; instead, the fields dissipate
shortly after the moment of impact in the form of outgoing waves.  This
situation is depicted in Fig.~\ref{fig:gauged-dissipate}.  For collision
velocities $v\gtrsim0.85$, we find that the majority of the field content
emerges along the axis of symmetry after the collision. However, the initial
\qballs{} are still difficult to distinguish in the aftermath as the field
magnitudes are greatly reduced and are also elongated in the radial direction.
This is accompanied by a spherical radiation pattern emanating from the origin.
An example of this scenario is depicted in App.~\ref{sec:suppfigs}
(Fig.~\ref{fig:gauged-tail2}). This lies in contrast to what is observed for
the logarithmic model where the dominant field components after the collision
take the form of gauged \qrings{} (cf.~Fig.~\ref{fig:gauged-Qrings}). However,
regardless of the final structure, we conclude that the equal-charge collisions
of solution PolyB can be considerably inelastic even at collision velocities
which are near-luminal.

Turning next to collisions of solution PolyB with a relative phase difference,
we find that charge transfer is once again the dominant outcome (as long as the
kinetic energy is sufficient to overcome the Coulomb repulsion). Similar to
what is observed for solution LogC, the \qballs{} created in this manner are
often unstable and may quickly fragment after the collision.  In some cases, we
even find that the instability can manifest via near-complete dispersal of the
fields so that the end result of the collision is just one remaining gauged
\qball{}.  An example of this behaviour for solution PolyB is given in
App.~\ref{sec:suppfigs} (Fig.~\ref{fig:gauged-transfer2}).  At the highest
velocities and for large phase differences, we find that the amount of charge
transfer is once again reduced.  For collisions of opposite charges, the
dynamics are generally independent of the relative phase with the main result
being the net annihilation of charge which is maximal at low collision
velocities. In contrast to what is observed for solution LogC
(cf.~Fig.~\ref{fig:gauged-EMpulse2}), we do not observe the formation of any
smaller \qballs{} during opposite-charge collisions involving solution PolyB.
Instead, the \qballs{} tend to continue propagating uniformly along the axis of
symmetry, though often being strongly perturbed by the annihilation process.

\section{\label{sec:conclusion} Conclusion}

In this work, we have performed high-resolution numerical simulations to study
head-on collisions of $U(1)$ gauged \qballs{}. Focusing on the relativistic
regime, we have studied the effects of various parameters (such as collision
velocity, relative phase, relative charge, and electromagnetic coupling strength) on
the outcome of the collision. Our simulations suggest that the outcome can
depend heavily on these parameters, resulting in dynamics which can be quite
distinct from those observed during collisions of ordinary (non-gauged)
\qballs{}.

We first examined the dynamics of gauged \qballs{} with small gauge coupling.
Here it was found that the dynamics for equal-charge collisions can generally
be divided into three regimes (the ``merger", ``fragmentation", and ``elastic"
regimes) depending on the collision velocity. We also studied the effect of
phase-difference and opposite-charge collisions, finding evidence for charge
transfer and annihilation, respectively. These findings are consistent with
what has been previously reported for ordinary (non-gauged) \qballs{}. Overall,
these results suggest that gauged \qballs{} with small gauge coupling can
behave like non-gauged \qballs{} during head-on collisions.

Turning to the case of large gauge coupling, we find that collisions of gauged
\qballs{} can lead to distinct dynamical behaviour due to the influence of the
electromagnetic field. For equal-charge collisions, the Coulomb force can cause
a repulsion which prevents the scalar field of each \qball{} from reaching a
state of significant interaction. This occurs at low collision velocities.  At
higher velocities, we find that collisions are rarely an elastic process;
instead, the main outcome is often a fragmentation of the colliding \qballs{} into
several smaller gauged \qballs{} or \qrings{}. This effect persists even at
collision velocities very close to the speed of light. Studying the effect of
phase difference on the collision outcome, we observe evidence for charge
transfer.  However, the gauged \qballs{} created during this process are often
unstable and tend to quickly break apart into smaller components.  For the case
of opposite-charge collisions, we find partial annihilation of the gauged
\qballs{} to be a generic outcome which can lead to the production of an
electromagnetic radiation pulse. Having studied these behaviours using both
polynomial and logarithmic scalar field potentials, we find that the collision
dynamics can differ slightly depending on the choice of potential. However, we
conclude that the main phenomena associated with gauged \qball{} collisions
(such as charge transfer, annihilation, and the inelasticity of the collisions)
are generally independent of the specifics of the model.

Since the present study has been limited to axisymmetry, it is 
interesting to ask how the dynamics may change in fully
three-dimensional simulations. This question will be addressed in a
future publication.
It would also be interesting to consider how quantum effects may
influence the dynamics of gauged \qballs{} similar to what has recently
been done for non-gauged \qballs{} \cite{Xie2024}. 
Finally, we comment that the results of this work could be extended by considering
more general scenarios in axisymmetry (such as collisions between
gauged \qballs{} with unequal $|Q|$) or by studying in further
detail the electromagnetic signal created
during the collisions. These scenarios may be relevant
for cosmological applications of gauged \qballs{} \cite{Hong2015,Hong2016,Hong2017,Jiang2024}.

\acknowledgments

We thank G.~Reid for providing valuable advice and feedback.  This work was
supported by the Natural Sciences and Engineering Research Council of Canada
(MWC, MPK), the Walter C.~Sumner Foundation (MPK) and the Province of British
Columbia (MPK).  Computing resources were provided by the Digital Research
Alliance of Canada and the University of British Columbia.\\

\appendix

\counterwithin{figure}{section}  

\section{Supplemental Figures}
\label{sec:suppfigs}

To supplement the figures presented in the main text, here we provide
additional plots which illustrate several interesting cases of gauged \qball{}
dynamics.

\bibliographystyle{apsrev4-2}
\bibliography{refs}  

\onecolumngrid 

\begin{figure}[p]
\includegraphics{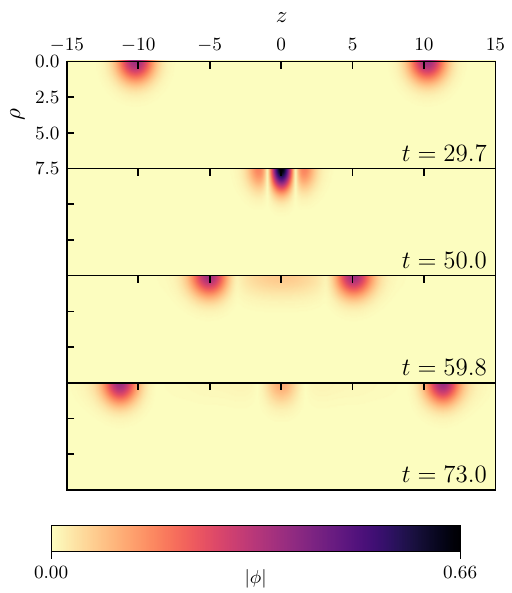}
  \caption{\label{fig:nongauged-fragment}
  Evolution of the scalar field modulus $|\phi|$ for a collision of solutions
  of type LogA with equal charge, velocity $v=0.5$, and phase difference
  $\alpha=0$.  The \qballs{} collide at $t\approx 50$ and pass through each
  other, leaving behind a smaller \qball{} remnant which remains perturbed at
  the origin.}
\end{figure}

\begin{figure}[p]
\includegraphics{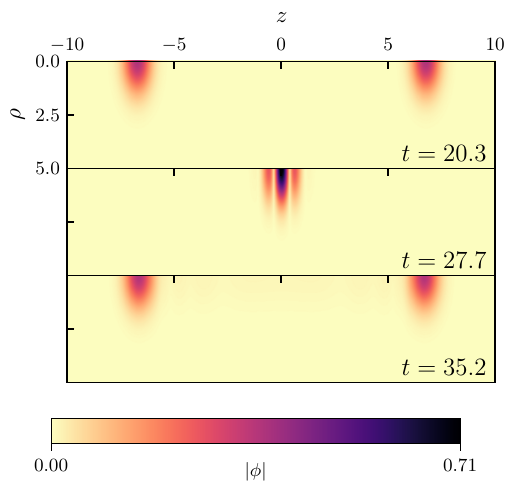}
  \caption{\label{fig:nongauged-passthru}
  Evolution of the scalar field modulus $|\phi|$ for a collision of
  solutions of type LogA with equal charge, velocity $v=0.9$, and phase
  difference $\alpha=0$. The \qballs{} collide at $t\approx 27$ and
  exhibit a destructive interference pattern.  After the collision, the
  \qballs{} emerge with profiles nearly identical to their initial
  state.}
\end{figure}

\begin{figure}[p]
\includegraphics{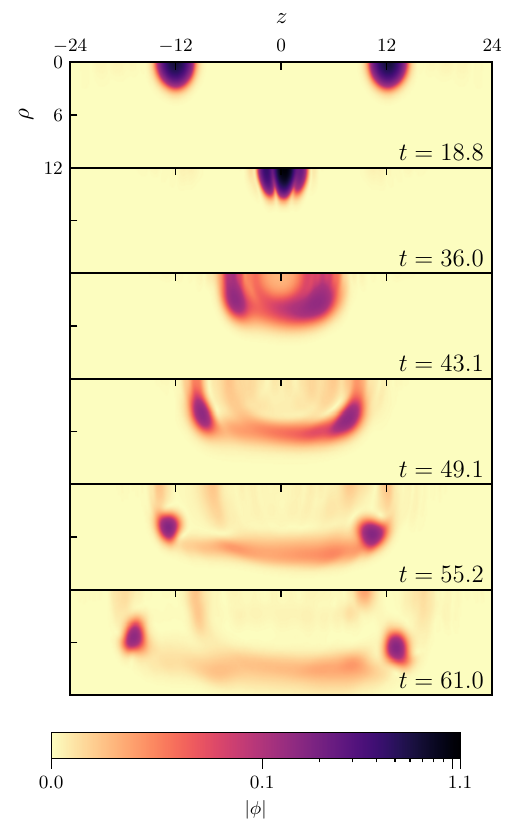}
  \caption{\label{fig:gauged-Qrings2}
  Evolution of the scalar field modulus $|\phi|$ for a collision of
  solutions of type LogC with equal charge, velocity $v=0.7$, and phase
  difference $\alpha=\pi/4$. The \qballs{} collide at $t\approx36$.
  After the collision, the field content predominantly takes the form of
  two \qrings{}. In this case, the phase difference manifests as an
  asymmetry in the dynamics about the plane $z=0$. Note that a hybrid
  colormap is used: field values below $|\phi|=0.1$ are mapped
  linearly to zero while values above this threshold are mapped
  logarithmically to the field maximum.}
\end{figure}

\begin{figure}[p]
\includegraphics{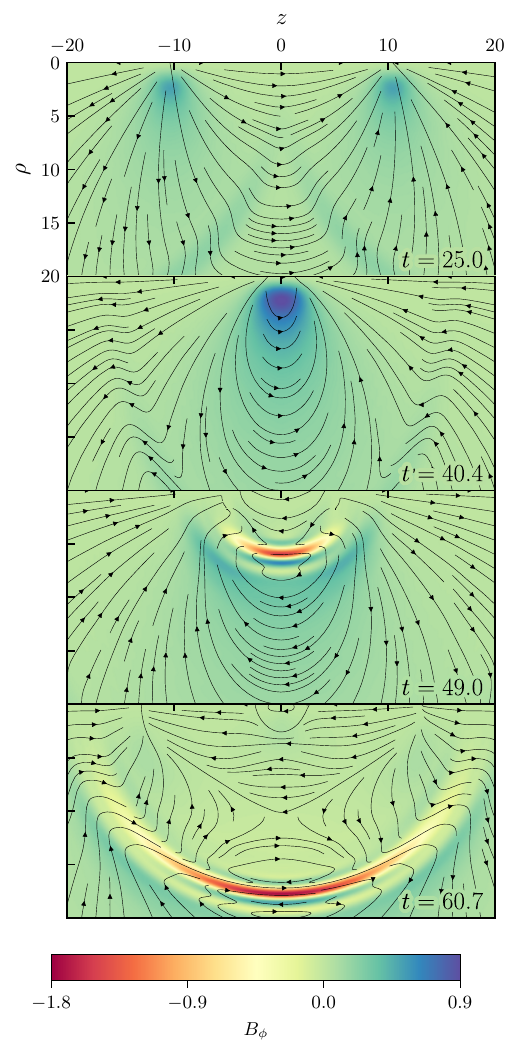}
  \caption{\label{fig:gauged-EMfield} Evolution of the electric field
  $\vec{E}$ and the magnetic field $\vec{B}$ for a collision of
  solutions of type LogC with opposite charge, velocity $v=0.6$, and
  phase difference $\alpha=0$.  The magnitude of the only non-zero
  component of the magnetic field, $B_\phi$, is represented using the
  colormap. The orientation of the electric field is represented using
  streamlines; the corresponding field magnitude is not reflected
  in the figure. The \qballs{} collide at $t\approx40$ and partially
  annihilate charge. After the collision, the fields resemble an
  outgoing wavefront. We note that the small-scale ``pulse" which is
  visible for $\rho\gtrsim 10$ in the first and second panel
  exists as a technical artefact of the gauged \qball{}
  initialization procedure at $z=\pm 25$.}
\end{figure}

\begin{figure}[p] 
\includegraphics{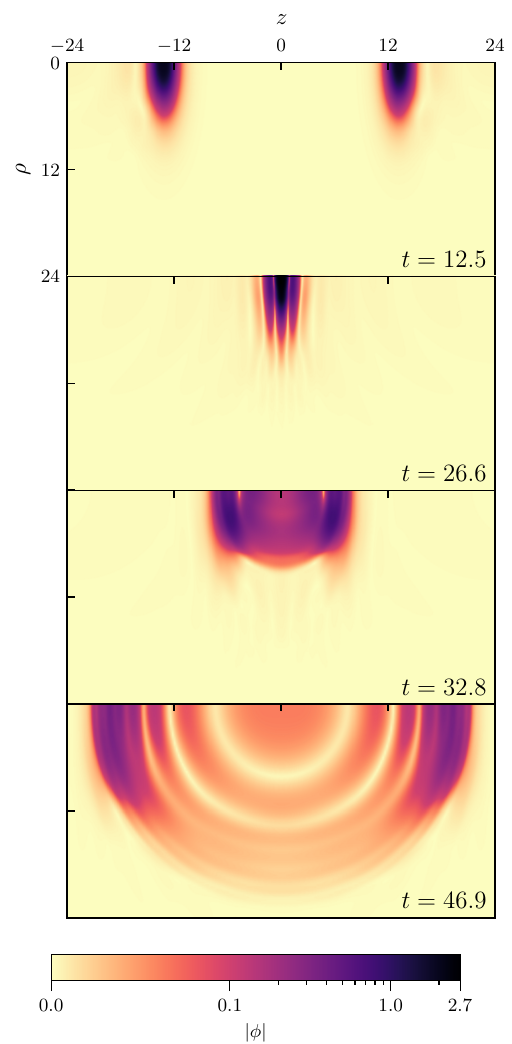}
  \caption{\label{fig:gauged-tail2}
  Evolution of the scalar field modulus $|\phi|$ for a collision of
  solutions of type PolyB with equal charge, velocity $v=0.95$, and
  phase difference $\alpha=0$.  The \qballs{} collide at $t\approx26$
  and form a destructive interference pattern.  After the collision, the
  majority of the field content continues travelling along the axis of
  symmetry and becomes elongated in the radial direction.  Note that a
  hybrid colormap is used: field values below $|\phi|=0.1$ are mapped
  linearly to zero while values above this threshold are mapped
  logarithmically to the field maximum.}
\end{figure}

\begin{figure}[p]
\includegraphics{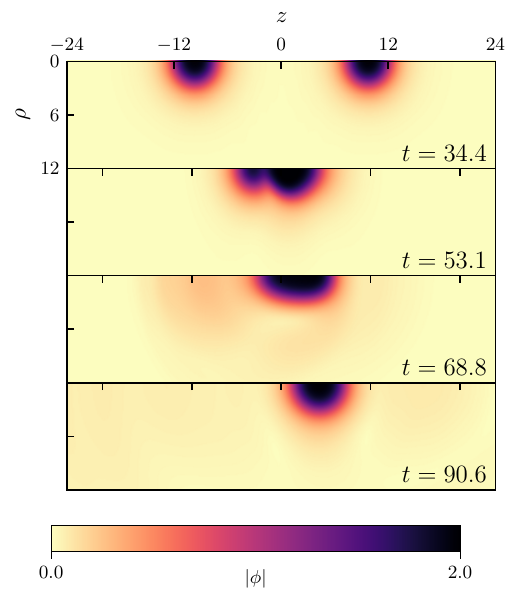}
  \caption{\label{fig:gauged-transfer2}
  Evolution of the scalar field modulus $|\phi|$ for a collision of
  solutions of type PolyB with equal charge, velocity $v=0.45$, and
  phase difference $\alpha=\pi/4$. The \qballs{} collide at $t\approx53$
  and transfer charge (as can be seen in the second panel).  After the
  collision, the smaller \qball{} created in this process quickly
  dissipates while the larger \qball{} travels slowly
  along the axis of symmetry.}
\end{figure}


\end{document}